\documentclass[aps,pra,floatfix,twocolumn,showpacs,preprintnumbers,amsmath,amssymb,a4paper,superscriptaddress]{revtex4-1}

\usepackage{graphicx}
\usepackage{xcolor}
\usepackage{wasysym}
\usepackage{amsfonts, amsmath, amssymb, bm, nccmath}
\usepackage{siunitx}
\usepackage{import}
\usepackage{color}
\usepackage[normalem]{ulem}
\usepackage{lpic}
\usepackage{textcomp}

\newcommand{\as}{a_{\mathrm{3D}}}

\begin{document}

\title{Confinement-induced resonances in ultracold atom-ion systems}

\author{V. S. Melezhik}
\email{melezhik@theor.jinr.ru}
\affiliation{Bogoliubov Laboratory of Theoretical Physics, Joint Institute for Nuclear Research, and Dubna State University,
Dubna, Moscow Region 141980, Russian Federation}

\author{A. Negretti}
\email{anegrett@physnet.uni-hamburg.de}
\affiliation{Zentrum
f\"{u}r Optische Quantentechnologien and The
Hamburg Centre for Ultrafast Imaging, Universit\"{a}t Hamburg,
Luruper Chaussee 149, 22761 Hamburg, Germany}

\date{\today}

\begin{abstract}
We investigate confinement-induced resonances in a system composed
by a tightly trapped ion and a moving atom in a waveguide. We
determine the conditions for the appearance of such resonances in
a broad region -- from the ``long-wavelength'' limit to the
opposite case when the typical length scale of the atom-ion
polarisation potential essentially exceeds the transverse waveguide width. We
find considerable dependence of the resonance position on the
atomic mass which, however, disappears in the ``long-wavelength and zero-energy''
limit, where the known result for the confined atom-atom scattering is
reproduced. We also derive an analytic and a semi-analytic formula
for the resonance position in the ``long-wavelength and zero-energy''
limit and we investigate numerically how the position of the resonance is
affected by a finite atomic colliding energy.
Our results, which can be investigated experimentally in the near future,
could be used to determine the atom-ion scattering length,
the temperature of the atomic ensemble in the presence of an ion impurity,
and to control the atom-phonon coupling in a linear ion crystal in interaction
with a quasi one dimensional atomic quantum gas.
\end{abstract}


\maketitle

\section{Introduction}
The interest in combining in the laboratory ultracold atoms and ions is increased considerably in the last few
years~\cite{Grier:2009,Zipkes:2010,Schmid:2010,Zipkes:2010b,Rellergert:2011,Hall:2012,Harter:2012a,Ratschbacher:2013,
Harter:2013,Haze:2013,Weckesser:2015,Kruekow:2016}. Their combination defines indeed a new quantum system characterised by an interaction
with different energy and length scales with respect to ultracold atoms, which allows to study the formation of molecular
ions~\cite{Cote:2002,Massignan:2005}, polarons~\cite{Casteels:2013}, density bubbles~\cite{Goold:2010}, mesoscopic
entanglement~\cite{Gerritsma:2012,Schurer:2016}, novel ground state properties~\cite{Schurer:2014} and collective excitations~\cite{Schurer:2015},
and quantum information processing~\cite{Doerk:2010,Joger:2014,Secker:2016}. Further, the interaction of the atoms with
the phonons of an ion crystal may serve to investigate solid-state phenomena~\cite{Bissbort:2013}.

Most of these fascinating theoretical proposals, however, assume that the atom-ion scattering length can be tuned in order to accomplish
the desired goal. Moreover, since the atom-phonon coupling along the symmetry axis of a linear ion chain interacting with a surrounding
quantum gas is very weak~\cite{Bissbort:2013}, the investigation of condensed-matter phenomena like Cooper-pairing in finite-size systems
of reduced dimensionality is prevented, but control over the atom-ion scattering length would help.

\begin{figure}[t!]
\includegraphics[scale=0.40,angle=0]{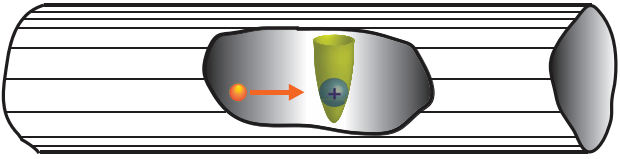}
\caption{ (Color online). Sketch of the ultracold collision between an atom confined in a waveguide-like trap
and a tightly trapped ion.}
\label{fig-sketch}
\end{figure}

Up to now, the atom-ion scattering length has been not yet experimentally measured, since the onset of s-wave scattering requires very low
temperatures (e.g., for Yb$^+$ and Li the required atom-ion collision energy is about 10 $k_B\times\,\mu$K). Thus, it is of paramount importance
to provide viable strategies to measure such scattering properties. Here we investigate the conditions for the appearance of confinement-induced
resonances (CIRs), which occur because of the tight confinement in one or two spatial dimensions, in atom-ion collisions confined by a waveguide-like
atomic trap, as illustrated in Fig.~\ref{fig-sketch}. Apart from its own interest,
CIRs in such combined system could provide an alternative method to experimentally measure the three-dimensional atom-ion scattering length,
the temperature of the quantum gas in the presence of an ion impurity as well as a way to enhance the atom-phonon coupling in a quasi-one dimensional
setting, enabling therefore the exploration of phonon-mediated interactions between atoms. Contrary to the atom-atom scenario, as we shall show,
in atom-ion systems the ratio between the interaction range and the width of the waveguide can be varied over a broader region allowing, for instance,
to observe an isotope-like effect in the CIR position. In the present study, however, we consider the scenario for which the ion is tightly confined (i.e, in
traps with frequencies $\omega\gtrsim 2\pi$ 100 kHz),
such that its motion can be discarded. Apart from simplifying the scattering problem, we underscore that such condition
can be fulfilled in current experiments (e.g., with optical traps~\cite{Huber:2014} or Paul traps, where frequencies
can even reach the MHz range~\cite{Leibfried:2003}).
Let us note that CIRs in atomic systems have attracted great interest, as they can be used to control the atom-atom
interaction~\cite{Guenter:2005,Haller:2010,Froehlich:2011}, e.g., for realising the so-called Tonks-Girardeau
gas~\cite{Kinoshita:2004,Paredes:2004} and the excited many-body phase known as the super-Tonks-Girardeau
gas~\cite{Haller:2009}.
Finally, theoretical investigations on atom-ion collisions have been carried out since long time now~\cite{Cote:2000,Zhang:2009,Idziaszek:2007,Idziaszek:2009,
Idziaszek:2011,Gao:2013,Tomza:2015}), and, in particular, quantum defect theory has been proven
to be a powerful tool for their description in the ultracold regime~\cite{Idziaszek:2007,Idziaszek:2009,
Idziaszek:2011,Gao:2013}, but the conditions for the appearance of CIRs were not yet considered.

The paper is structured as follows: In Sec.~\ref{sec:pb} we describe the physical problem we are interested in by providing the system Hamiltonian, the relevant energy and length scales
of the problem, and important details of the procedure we employed in order to determine the scattering amplitudes. In Sec.~\ref{sec:atom-atom} we briefly describe the known results of
CIRs in ultracold atom-atom collisions. This will help the understanding of the energy-dependent pseudopotential method we shall employ in the atom-ion scenario in the ``long-wavelength"
limit (LWL). In Sec.~\ref{sec:atom-ion} we investigate CIRs in atom-ion systems in the LWL limit and beyond. In Sec.~\ref{sec:heating} we shall have a general discussion on our
findings, especially with respect to current experimental limitations as micromotion and finite temperature. Finally, in Sec.~\ref{sec:end} we shall draw our conclusions and provide an outlook.

\section{Problem formulation}
\label{sec:pb}

Let us consider an atom of mass $m_A$ experiencing the external harmonic potential $m_A\omega_{\perp}^2(x_A^2+y_A^2)/2$ and an ion of mass $m_I$ in the harmonic trap
$m_I\omega^2\vert{\bf r}_I\vert^2/2$. The quantum dynamics of this combined system is described by the following Hamiltonian

\begin{align}
\hat{H} & =-\frac{\hbar^2}{2m_A}\nabla_A^2 +
\frac{1}{2}m_A\omega_{\perp}^2(x_A^2+y_A^2) \nonumber\\
\phantom{=}& -\frac{\hbar^2}{2m_I}\nabla_I^2
+\frac{1}{2}m_I\omega^2\vert{\bf r}_I\vert^2 + V(\vert{\bf r}_A-{\bf
r}_I\vert),
\label{hamiltonian}
\end{align}
where the terms indicated by the indexes $A$ and $I$ describe the atomic and ionic parts of the Hamiltonian correspondingly, and ${\bf r}_A\equiv (x_A,y_A,z_A)$ and ${\bf r}_I\equiv (x_I,y_I,z_I)$
are the atomic and ionic coordinates. The interaction between the two particles is described by the long-range potential

\begin{equation}
V(\vert{\bf r}_A-{\bf r}_I\vert)
\rightarrow  - \frac{C_4}{\vert{\bf r}_A-{\bf r}_I
\vert^4}
\label{interaction}
\end{equation}
as $\vert{\bf r}_A-{\bf r}_I \vert \rightarrow \infty$. Here $C_4=\alpha_p e^2/(8 \pi\epsilon_0)$ with $\alpha_p$
being the static dipolar polarisability of the atom, $e$ the electron charge, and $\epsilon_0$ the vacuum
permittivity. Note that the distance at which the atom-ion interaction starts to deviate from Eq.~(\ref{interaction})
is typically given by the size of the atom-ion complex core (a few tens of Bohr radii),
which is much smaller than any length scale in our scattering problem. Furthermore, the interaction~(\ref{interaction})
is characterised by the length scale $R^*=\sqrt{2\mu C_4}/\hbar$ and by the energy scale $E^*=\hbar^2/[{2}\mu (R^*)^2]$,
where $\mu$ denotes the reduced mass.

Hereafter we neglect the kinetic energy operator $-\hbar^2/(2m_I)\nabla_I^2$ for the ion, as the ion is assumed to be strongly confined.
This assumption is even more well justified when $m_I \gg m_A$, like for the atom-ion pair Li-Yb$^+$. Assuming that the (static) ion is located
at ${\bf r}_I=0$, the Schr\"odinger equation for the atom is given by

\begin{equation}
\left(-\frac{1}{m_A}\nabla^2_r
+m_A \omega_{\perp}^2\rho^2 + \frac{C_{12}}{r^{12}}-\frac{1}{r^4}\right)\psi({\bf
r}) = E \psi({\bf r})
\label{2Dhamiltonian}
\end{equation}
with the boundary condition for $z \rightarrow \pm\infty$

\begin{equation}
\psi(z,\rho) = [\exp(ikz) + f^{\pm}(k,a_{\perp},\as)\exp(i k\vert z
\vert)]\varphi_0(\rho),
\label{asymptotics}
\end{equation}
whereas in the transverse direction the atomic wavefunction is assumed to be in the ground state $\varphi_0(\rho)$ of the two-dimensional (2D)
harmonic oscillator. Here $f^{\pm}$ denote the scattering amplitudes at $z \rightarrow \pm\infty$, the lengths are given in $R^*$ units, $m_A$ in
units of $\mu$, $\omega_{\perp}$ in units of $\omega^*=2 E^*/\hbar=\hbar/[\mu (R^{*})^2]$, i.e. $\omega_{\perp}/\omega^* = (R^*/a_{\perp})^2$
with $a_{\perp}=\sqrt{\hbar/(\mu\omega_{\perp})}$ the width of the transverse harmonic trap, and ${\bf r} \equiv (x,y,z)$, $r=\vert\mathbf{r}\vert$, ${\bf \rho} = (x,y)$.
Besides, the detalyze atom-ion interaction $C_{12}/r^{12}-C_4/r^4$ gives the correct decay of the interaction for $r\rightarrow\infty$, whose effective
range is on the order of $R^*$, and a realistic number of bound states. Hence, the three-dimensional scattering problem for the atom with a fixed
scattering centre given by the atom-ion polarisation potential is reduced to the integration of the two-dimensional Schr\"odinger equation~(\ref{2Dhamiltonian})
at a fixed total energy $E = E_{\parallel} + \hbar\omega_{\perp}$ and subsequent extraction of $f^{\pm}(k,a_{\perp},\as)$ at the fixed longitudinal
wave number $k=\sqrt{m_A E_{\parallel}}/\hbar$ by comparing (i.e., fitting) the numerically obtained solution of Eq.~(\ref{2Dhamiltonian}) at
$\vert z\vert\rightarrow\infty$ with the asymptotic wavefunction given by Eq.~(\ref{asymptotics}). Here the wave number $k$ is defined
by the longitudinal colliding energy $E_{\parallel}=E-\hbar\omega_{\perp}$, where $\hbar\omega_{\perp}$ is the ground state energy
of the 2D harmonic oscillator. To this end, we employ the computational scheme described in Ref.~\cite{Melezhik:2016}, which has been successfully
applied to the atom-atom~\cite{Melezhik:2003,Saeidian:2008,Saeidian:2012} as well as to the dipole-dipole confined scattering~\cite{Giannakeas:2013,Koval:2014}.

\section{Atom-atom confinement-induced resonances}
\label{sec:atom-atom}

Let us briefly recall the main results of the confined atom-atom scattering in the LWL, i.e., $E\ll\hbar^2/[{2} \mu (r^{*})^2]$~\footnote{
Here the effective range of the atom-atom interaction is defined as $r^{*2} =\sqrt{2\mu C_6}/\hbar$ with $C_6$ being the dispersion coefficient.}, obtained
within the pseudopotential approximation, as they will be useful when we shall discuss our findings in the atom-ion setting in that limit (i.e., $R^* \ll a_{\perp}$).

In Refs.~\cite{Olshanii:1998,Moore:2001} the position of the CIR was defined as the pole of the quasi-1D coupling constant
$g_{\mathrm{1D}}=\lim_{k\rightarrow 0} \Re[f_g(k,a_{\perp}/\as)]/\Im[f_g((k,a_{\perp}/\as)]\hbar^2 k/\mu \rightarrow \pm\infty$, where $f_g=(f^{+}+f^{-})/2$,
fulfilling the condition

\begin{equation}
\frac{a_{\perp}}{\as}= -\zeta(1/2) = 1.4603\dots
\label{CIRaa}
\end{equation}
Here $\as=-\lim_{k\rightarrow 0} f_0(k)$ is the 3D s-wave (atom-atom) scattering length in free-space, which is defined by the zero-energy limit of the s-wave
scattering amplitude $f_0(k)$ with $\zeta(x)$ being the Hurvitz Zeta function. Note that in this limit we have $\Re[f_g(k,a_{\perp}/\as)]\rightarrow -1$,
$\Im[f_g(k,a_{\perp}/\as)]\rightarrow 0$, and the transmission coefficient $T(k,a_{\perp}/\as)= \vert 1+ f^{+}(k,a_{\perp}/\as)\vert^2$ goes to
zero~\cite{Kim:2006,Saeidian:2008,Saeidian:2012}, that is, we have total reflection of the colliding particles. Moreover, the condition (\ref{CIRaa}) defining the appearance of CIR
does not have any reliance on the atomic mass. In a series of works~\cite{Kim:2006,Saeidian:2008,Saeidian:2012}, the condition~(\ref{CIRaa}) was confirmed by
numerically computing  the poles of the coupling constant $g_{\mathrm{1D}}(k,a_{\perp}/\as)$ at $k\rightarrow 0$. Interestingly, it was recently shown that
Eq.~(\ref{CIRaa}) is still applicable for CIRs of the confined dipole-dipole scattering in the week dipole-dipole interaction limit and $k\rightarrow 0$~\cite{Giannakeas:2013},
whose interaction scales as $1/r^3$, which decays even slower than the atom-ion interaction~(\ref{interaction}). However, in the first fully numerical analysis
reported in Ref.~\cite{Bergeman:2003} of the confined atom-atom scattering near the CIR at a small, but fixed, wave number $k=\sqrt{m_AE_{\parallel}}/\hbar$, it was
observed a difference between the numerically estimated value of $a_{\perp}/\as$ at the point of CIR and the analytic formula~(\ref{CIRaa}). The difference was not ascribable to
numerical imprecision, but rather because of the non-zero colliding energy. In a subsequent study~\cite{Naidon:2007}, it has been showed that by using the
formula~(\ref{CIRaa}) with the replacement $\as \rightarrow \as (k)$ in the so-called effective-range approximation~\cite{Blume:2002,Bolda:2002}

\begin{equation}
\frac{1}{\as(k)} = -k\cot \delta_0(k) = \frac{1}{\as}-\frac{1}{2} R_0
k^2 + ...
\label{effectiver0}
\end{equation}
the aforementioned difference can still be well described by the pseudopotential approach ($R_0$ is the so-called effective-range parameter). There the wave number
$k=\sqrt{m_AE}/\hbar$ was defined by identifying the longitudinal atomic energy $E_{\parallel}$ in a waveguide with the colliding energy in free-space $E$, that is, $E=E_{\parallel}$.

\section{Atom-ion confinement-induced resonances}
\label{sec:atom-ion}

In this section we shall investigate CIRs in the atom-ion scenario. We shall first consider the ``long-wavelength" limit,
that is, $R^*\ll a_{\perp}$, and then the case $R^*\gtrsim a_{\perp}$, but always under the ``static ion" assumption.

\subsection{The long-wave length limit: $R^*\ll a_{\perp}$}

Given the aforementioned studies on atom-atom CIRs, especially the ones based on the energy-dependent scattering length, it seems natural also for the confined
atom-ion scattering to look for CIRs in the region near to the point defined by Eq.~(\ref{CIRaa}), at least in the LWL. In the present case, however, because of the
long-range character of the atom-ion interaction~(\ref{interaction}), we have to add the linear, $\sim k$, and the logarithmic, $\sim \ln(k)k^2$, terms in the expansion~(\ref{effectiver0}).
To obtain an expression analogue to Eq.~(\ref{CIRaa}) describing the position of the atom-ion CIR in the LWL including all the terms up to $k^2$, it is needed not only to
use the energy-dependent scattering length $\as(k)$ on the left-hand-side of Eq.~(\ref{CIRaa}), but also to include the corresponding energy-dependent terms on its right-hand-side.
To this aim, we use the following expression for the scattering amplitude

\begin{align}
f_g(k,a_{\perp}/\as) =
-\frac{2}{2-ia_{\perp}k[\frac{a_{\perp}}{\as(k)}+\zeta(\frac{1}{2})+\frac{1}{8}\zeta(\frac{3}{2})a_{\perp}^2k^2]},
\label{amplitude}
\end{align}
which was derived in the LWL in Ref.~\cite{Moore:2001} for the confined atom-atom scattering, but for constant $\as$. By using this expression and the effective-range approximation
for the energy-dependent scattering length $\as(k)$, including all the terms up to $k^2$, we arrive at the following equation

\begin{align}
\frac{a_{\perp}}{\as(k)}  =
-\zeta\left(\frac{1}{2}\right)-\frac{1}{8}\zeta\left(\frac{3}{2}\right)(a_{\perp}k)^2 = \nonumber\\
 = 1.4603 - 0.6531 \left(\frac{m_A}{\mu}\right) \left(\frac{E_{\parallel}}{\hbar\omega_{\perp}}\right).
\label{CIRai}
\end{align}
Equation~(\ref{CIRai}) defines the position of the CIR for the atom-ion confined scattering case
at non-zero longitudinal energy within the ``static ion'' and the LWL approximations, and it is one of our main results.
It follows from the effective-range expansion for $\as(k)$ describing the low-energy atom-ion scattering in
free-space~\cite{OMalley:1961,Levy:1963} that the left-hand-side of Eq.~(\ref{CIRai}) can be represented by

\begin{align}
\frac{a_{\perp}}{\as(k)} & = a_{\perp}\left\{\frac{1}{\as} - \frac{\pi
}{3(\as)^2}k - \frac{4}{3\as}\ln\left(\frac{k}{4}\right)k^2
\right.\nonumber\\
\phantom{=}& \left. - \frac{1}{2}R_0^2k^2-
\left[\frac{\pi}{3}+\frac{20}{9\as}-\frac{\pi}{3(\as)^2}-\frac{\pi^2}{9(\as)^3} \right.\right.\nonumber\\
\phantom{=}&-\left.\left.\frac{8}{3\as}\psi'\left(\frac{3}{2}\right)\right]k^2
\right\},
\label{effectiver02}
\end{align}
where $\psi'(\frac{3}{2})=0.82072...$
denotes the digamma function. It indicates that the CIR position depends on  the atom-ion scattering length $\as$, the effective-range
$R_0$ in free-space, the width of the confining trap $a_{\perp}$, and the static dipolar polarisability of the atom $\alpha_p$.
The reliance on the latter is due to the fact that all lengths and $k$-vectors in Eq.~(\ref{effectiver02}) are given in units of $R^*$ and $R^{*-1}$.

\begin{figure}[t!]
\hspace*{4mm}\includegraphics[scale=0.3]{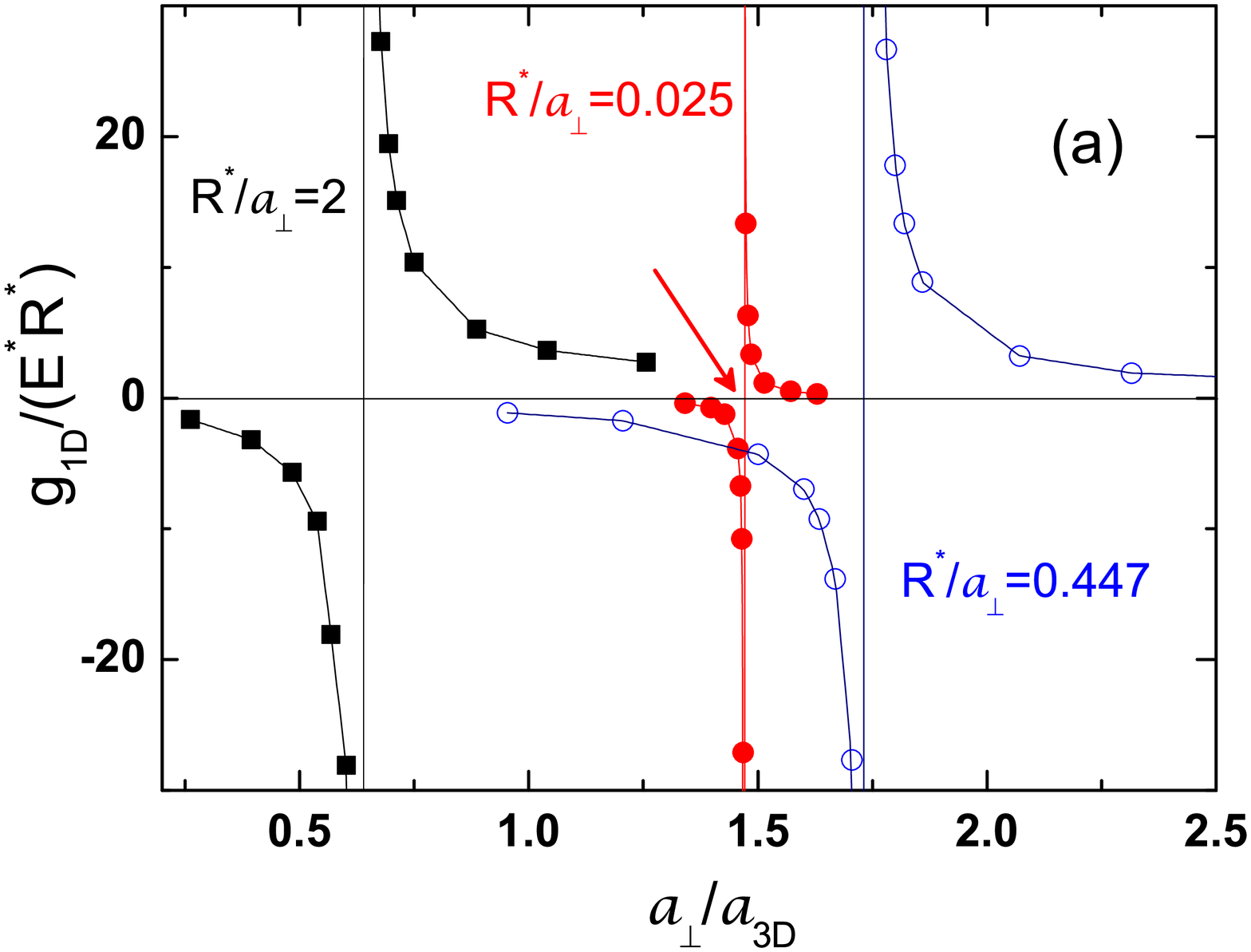}
\includegraphics[scale=0.3]{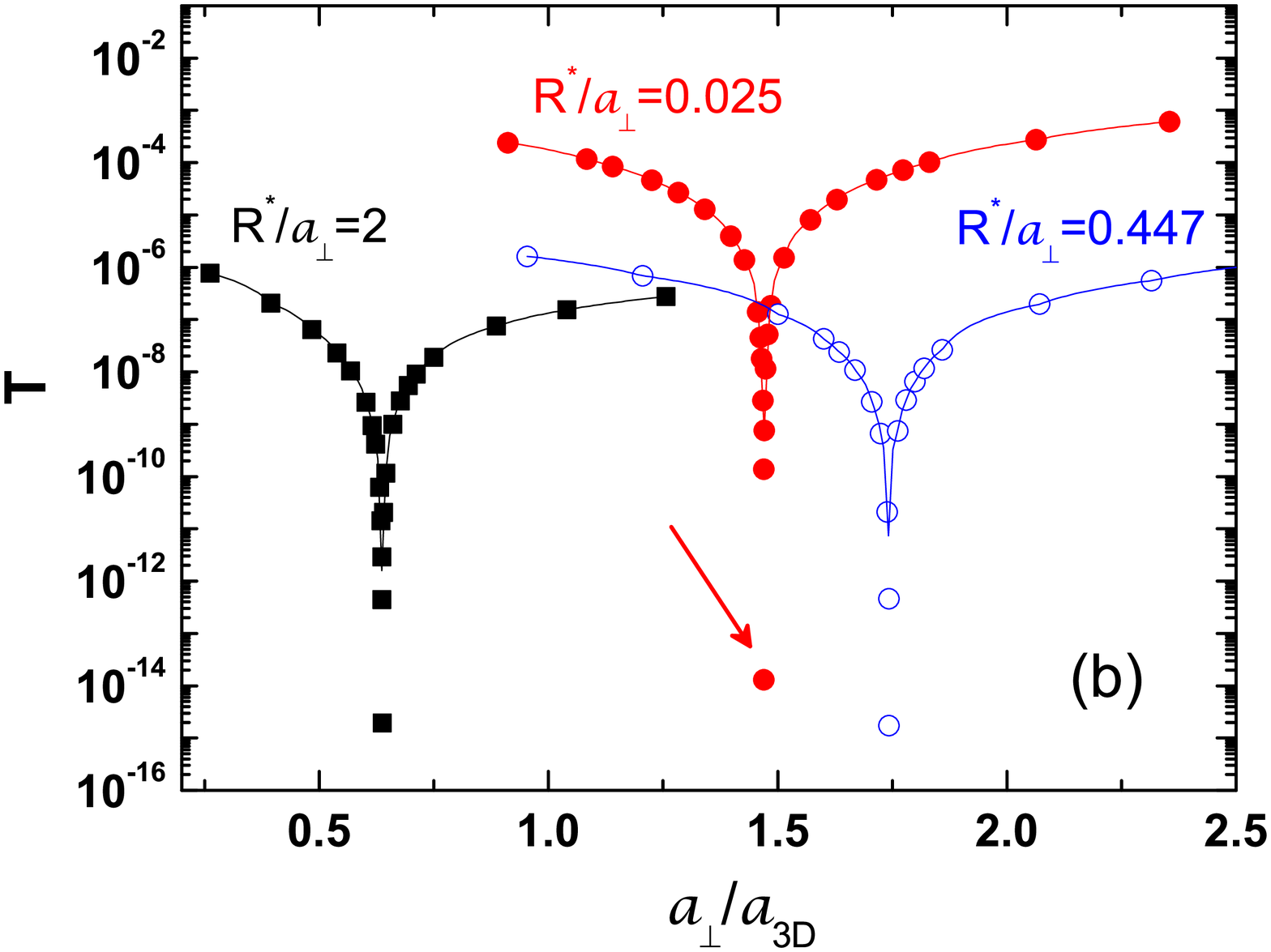}
\caption{(Color online). Effective coupling constant $g_{\mathrm{1D}}$ (a) and transmission $T$ (b) as a function of
$\as^{-1}$ numerically calculated for the atom-ion pair $^{6}$Li-$^{171}$Yb$^+$ at the
zero-energy limit $E_{\parallel}/E^*=10^{-6}$
for three different values of $a_{\perp}$. The
solid lines are a guide to the eye and the arrows indicate the CIR position given by Eq.~(\ref{CIRaa}).}
\label{fig2}
\end{figure}

In the numerical integration of the 2D Schr\"odinger equation~(\ref{2Dhamiltonian}) we have used Eq.~(\ref{CIRai}) as a starting point in the
search for atom-ion CIRs. In particular, we have followed this procedure: Firstly, for fixed $E/E^*\ll 1$ and $\omega_{\perp}$ (i.e., fixed
$a_{\perp}$), we have solved Eq.~(\ref{2Dhamiltonian}) by varying the coefficient $C_{12}$ in order to determine the position of the CIR
for a certain atom-ion pair (i.e., chosen $C_4$ coefficient, that is, a fixed value of $R^*$). We note that vary the coefficient $C_{12}$
for a fixed value of $E/E^*$ and $a_{\perp}$ is equivalent to vary the three-dimensional atom-ion scattering length $a_{\mathrm{3D}}$, which can
be attained by means of magnetic Feshbach resonances (see Refs.~\cite{Idziaszek:2009,Tomza:2015} for studies on such resonances in the
atom-ion context). From the obtained solution $\psi(\mathbf{r})$ we extract the scattering amplitude $f^{\pm}(k,a_{\perp}/\as)$ by fitting the
numerically obtained solution for $\mid z\mid\gg1$ with the asymptotic~(\ref{asymptotics}). With this knowledge, the coupling constant
$g_{\mathrm{1D}}$ is calculated. Once the coefficient $C_{12}$ for which $g_{\mathrm{1D}}$ has a pole is found, we solved Eq.~(\ref{2Dhamiltonian})
once again with that $C_{12}$ at $E=E_{\parallel} (k=\sqrt{m_A E_{\parallel}})$ for $\omega_{\perp} = 0$, that is, in free space, such that the 3D
s-wave scattering amplitude $f_0(k)$ is determined, and therefore $\as(k)=-f_0(k)$. Figure~\ref{fig2} shows the coupling constant $g_{\mathrm{1D}}$
(upper panel) as well as the transmission coefficient $T$ (lower panel) as a function of $\as^{-1}$ calculated close to the CIR in the
(almost) zero-energy limit for $R^*$ corresponding to the pair $^{6}$Li-$^{171}$Yb$^+$ and for three different values of $\omega_{\perp}$
(i.e., $a_{\perp}$). In particular, it demonstrates that the numerically calculated CIR position for $R^*= 0.025 a_{\perp} \ll a_{\perp}$ is in good
agreement with Eq.~(\ref{CIRaa}) (see the arrow close to the red full circles). This result shows that by looking for the CIR position one can infer the
value of the atom-ion scattering length $a_{\mathrm{3D}}$. Indeed, at the CIR position and in the zero-energy limit we estimate $\as$ to be of the order
$10^2$ nm for the atom-ion pair $^{6}$Li-$^{171}$Yb$^+$ for currently used atomic trap frequencies $\omega_{\perp}=2\pi\times (100-10)$kHz,
which is within the range of values estimated by Tomza et al.~\cite{Tomza:2015}.

\begin{figure}[t!]
\includegraphics[scale=0.35]{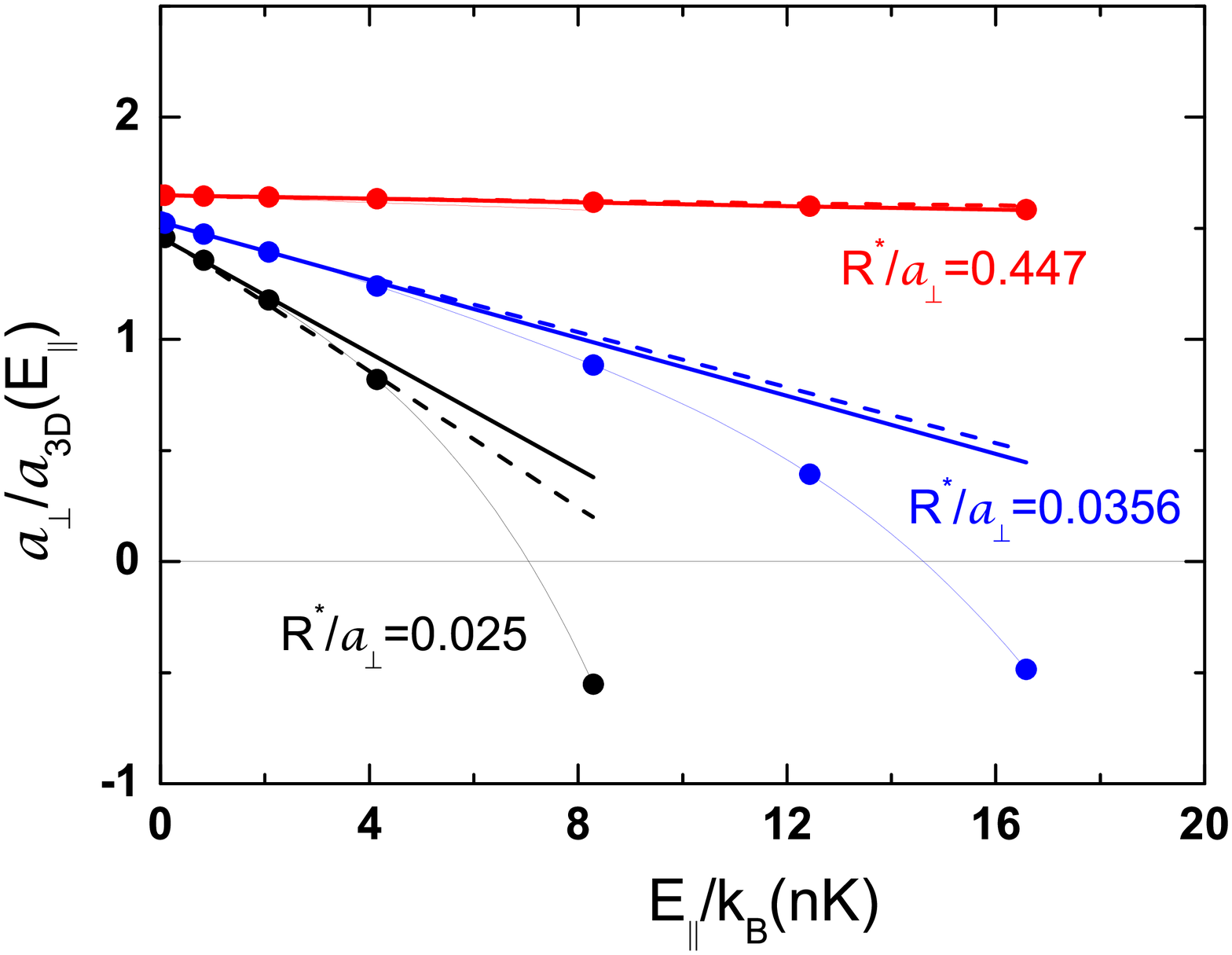}
\caption{(Color online). Ratio $a_{\perp}/\as(E_{\parallel})$ calculated at the location of the atom-ion CIR position as a function of $E_{\parallel}$ for
three different values of the trap frequency $\omega_{\perp}=(R^*/a_{\perp})^2$ for the atom-ion pair $^{6}$Li-$^{171}$Yb$^+$. The circles
represent the calculated values of $a_{\perp}/\as(k)$ via the integration of Eq.~(\ref{2Dhamiltonian}) with the boundary condition~(\ref{asymptotics}).
The solid curves correspond to Eq.~(\ref{CIRai}) at $R^*/a_{\perp}=0.025$ and Eq.~(\ref{CIRaic}) at higher values of $R^*/a_{\perp}$, whereas
the dashed curves correspond to Eq.~(\ref{effectiver02}) with the $\as$ and $R_0$ extracted from the curve $1/\as(k)$ calculated at the zero-energy
limit in free-space, but with the parameter $C_{12}$ in the atom-ion interaction~(\ref{interaction}) fixed at the CIR position.}
\label{fig3}
\end{figure}

A more detailed analysis is given in Fig.~\ref{fig3}, where again the atom-ion CIR position in the zero-energy limit ($k\rightarrow0$) is defined
by Eq.~(\ref{CIRaa}). Moreover, Eqs.~(\ref{CIRai}) and~(\ref{effectiver02}) well describe the energy dependence of the calculated CIR position
$a_{\perp}/\as(k)$ at low energies in the case $R^*/a_{\perp}=0.025$. With increasing $k$, however, the calculated curve $a_{\perp}/\as(k)$ of the CIR
begins to significantly deviate from the analytical curve~(\ref{CIRai}) at $E_{\parallel} > 3$ nK and from~(\ref{effectiver02}) at
$E_{\parallel} >5$ nK. With increasing $R^*/a_{\perp}=\sqrt{\omega_{\perp}/\omega^*}$ we have found a considerable shift
$\Delta=\Delta(R^*/a_{\perp})$ of the CIR position from the point defined by Eq.~(\ref{CIRai}) according to:

\begin{align}
\frac{a_{\perp}}{\as(k)} =
1.4603 +\Delta\left(\frac{R^*}{a_{\perp}}\right)-0.6531 \left(a_{\perp}k\right)^2=\nonumber \\
=1.4603 +\Delta\left(\frac{R^*}{a_{\perp}}\right)-0.32655 \left(\frac{m_A}{\mu}\right)\left(\frac{a_{\perp}}{R^*}\right)^2\left(\frac{E_{\parallel}}{E^*}\right).
\label{CIRaic}
\end{align}
We note that the shift $\Delta(R^*/a_{\perp})$ was not included in the formulas derived in Ref.~\cite{Moore:2001} [see Eq.~(\ref{amplitude}) therein].
Figure~\ref{fig3} nicely illustrates the enhanced range of applicability of Eq.~(\ref{CIRaic}) as compared with Eq.~(\ref{CIRai}). Moreover, the range of
applicability of Eq.~(\ref{CIRaic}) even extends with increasing $R^*/a_{\perp}$ due to the fact that the last term in Eq.~(\ref{CIRaic}) decreases.
We underline that the last term in Eq.~(\ref{CIRaic}), linearly depending on energy, coincides with the last term in Eq.~(\ref{CIRai}), which has been
obtained in the LWL approximation. 
 Note that the case $R^*/a_{\perp}=0.447$ ($\omega_{\perp}\simeq 2\pi\times 71$ kHz) falls already into the region of experimentally reachable values of atomic traps \cite{Harter:2013,Cetina:2012}. This clearly shows that already moderate trap frequencies allow to explore a broader range of $R^*/a_{\perp}$, which is not the case for atom-atom collisions in a waveguide,
where one has either to rely on Feshbach resonances or produce very tight traps.

\begin{figure}[t!]
\includegraphics[scale=0.35]{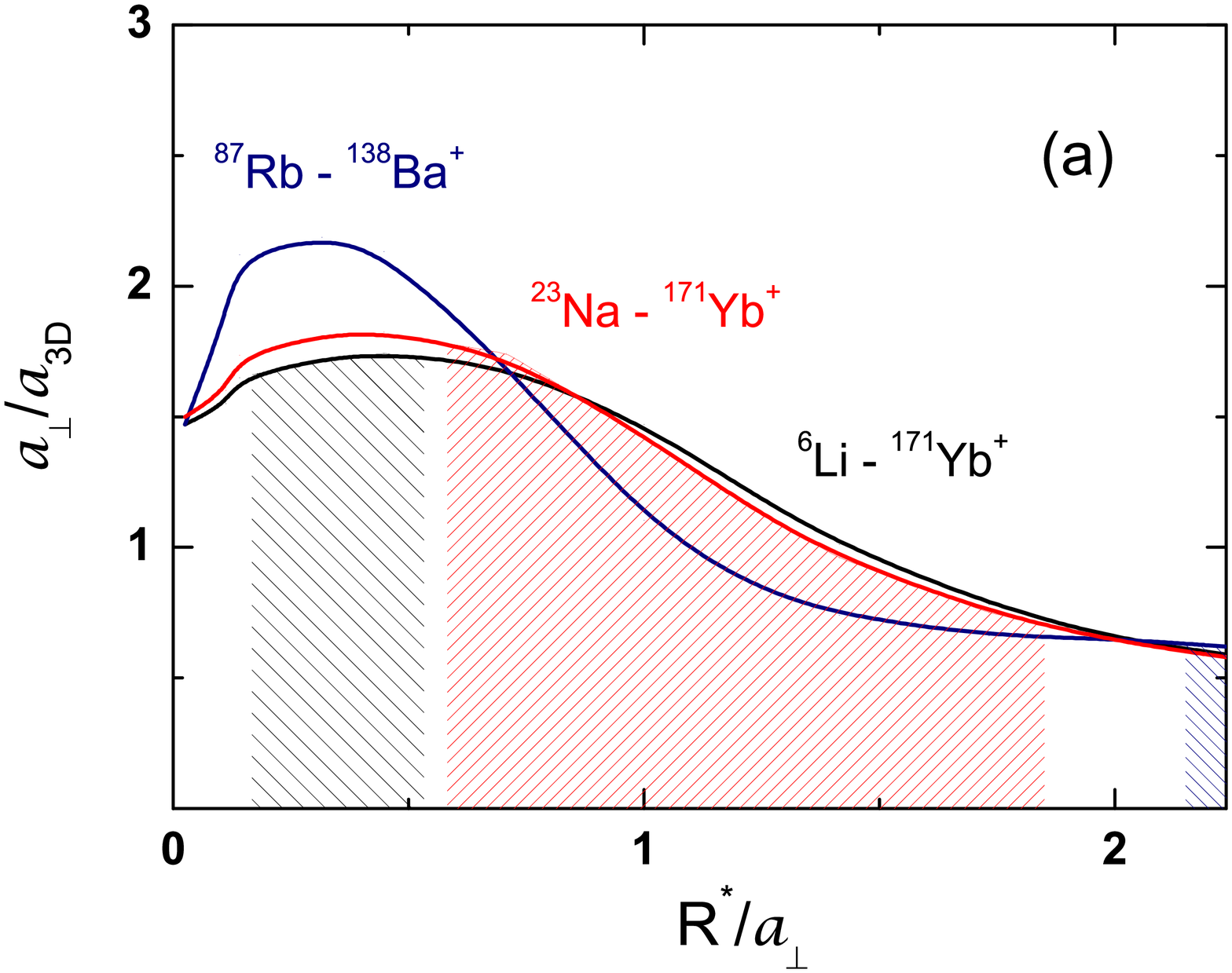}
\includegraphics[scale=0.35]{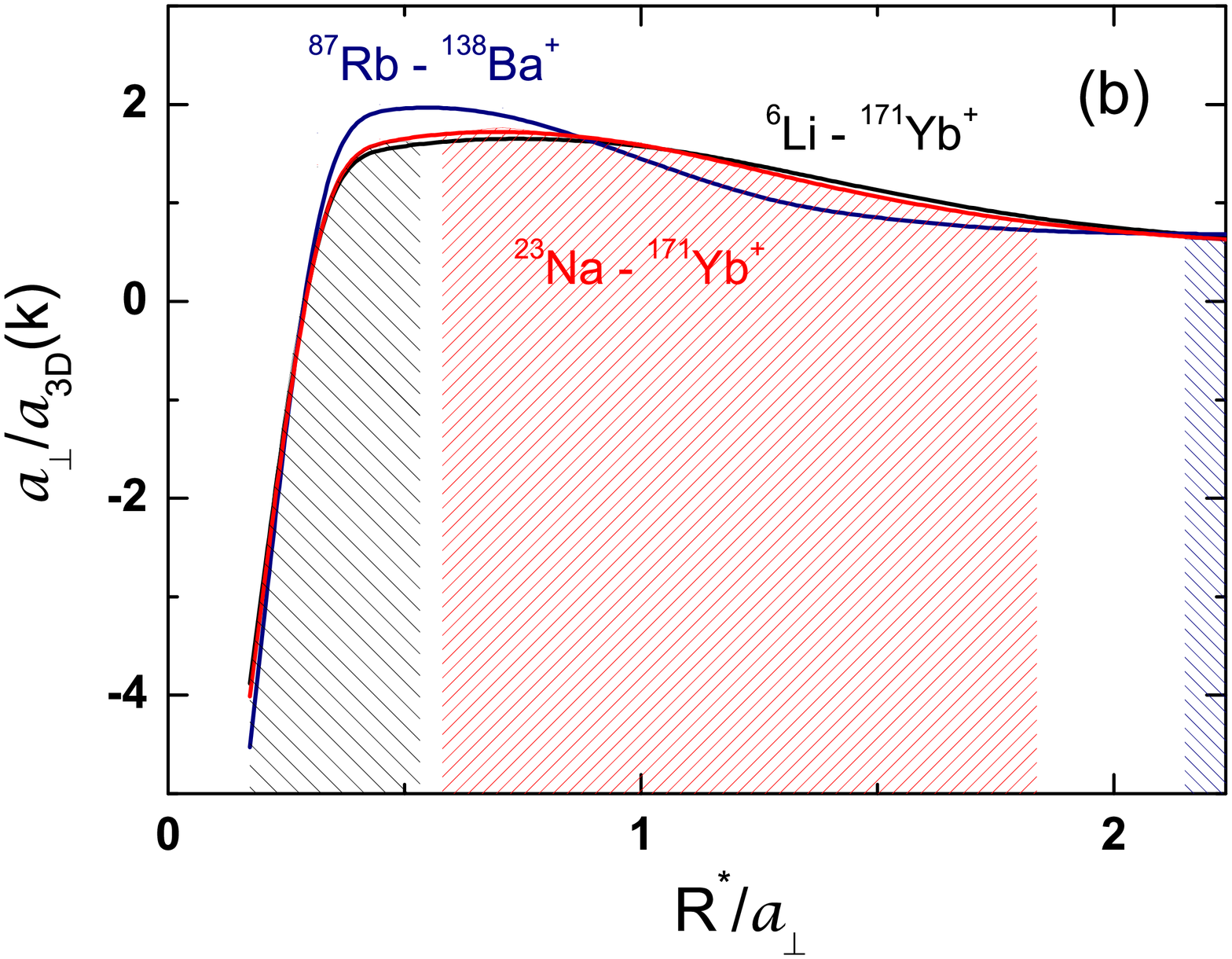}
\caption{ (Color online). (a) The ratio $a_{\perp}/\as=1.4603+\Delta(m_A/\mu,R^*/a_{\perp})$ in the points of the CIR as a function of $R^*/a_{\perp}$ calculated for different
atom-ion pairs at the zero-energy limit $E_{\parallel}/E^*=10^{-6}$. (b) The ratio $a_{\perp}/\as(k)$ in the points of the CIR as a function of $R^*/a_{\perp}$ calculated for
different atom-ion pairs at thefinite colliding energy $E_{\parallel}/E^*=0.117$, which corresponds to $E_{\parallel}/k_B=1\,\mu$K for the pair $^{6}$Li-$^{171}$Yb$^+$,
$6$nK for  $^{87}$Rb-$^{138}$Ba$^+$, and $80$nK for  $^{23}$Rb-$^{171}$Yb$^+$. Shaded areas indicate the range $\omega_{\perp}=2\pi\times (10-100)$kHz for the
atomic transverse trap frequencies, which are reachable in current experiments.}
\label{fig4}
\end{figure}

Let us conclude this section with a few remarks. We derived formulas for the atom-ion CIR position~(\ref{CIRai}) and~(\ref{CIRaic}) for a fixed atomic wave-number
$k=\sqrt{m_AE_{\parallel}}/\hbar$ and numerically investigated their applicability in the region $0\leq E_{\parallel}\leq 2\hbar\omega_{\perp}$, namely up to the threshold
for transverse atomic excitation. We note that these formulae are also applicable for atom-atom CIRs, where the inter-particle interaction is even more short-range.
Moreover, the atom-atom CIRs for colliding energies in the range $0\leq E_{\parallel}\leq 6\hbar\omega_{\perp}$, that is, with excitations up to third transverse mode, were
investigated numerically in Ref.~\cite{Saeidian:2008}. Up until now, however, to the best of our knowledge, such corresponding analytic formulae for describing the dependence
of the CIR position on the energy were not suggested.

Finally, as we already mentioned in the introduction, we argue that by means of CIRs the atom-phonon coupling can be controlled and enhanced in a quasi-1D setting,
similarly to the scenario described in Ref.~\cite{Bissbort:2013}, where electrons in a 1D solid crystal were replaced with neutral (fermionic) atoms. In that work it was shown
that, while the transverse atom-phonon coupling was on the order of $E^*$, the longitudinal coupling, namely along the ion crystal, was several order of magnitude smaller (i.e.,
$\sim 3\times 10^{-5} E^*$). As it was shown there, the atom-phonon coupling is proportional to

\begin{align}
\int_{\mathbb{R}}dz \phi_q^*(z) V^{\prime}(z) \phi_{q^{\prime}}(z),
\end{align}
where $\phi_q(z)$ is the Bloch function of quasi-momentum $q$, and $V^{\prime}(z)$ is the derivative with respect to $z$ of the atom-ion polarisation potential. As it has
been shown later in Ref.~\cite{Negretti:2014}, for ion separations larger than $5 R^*$, as in Ref.~\cite{Bissbort:2013}, the atom-ion polarisation potential is well described
by the pseudopotential with 1D energy-dependent scattering lengths. This implies that the atom-phonon coupling is also proportional to the quasi-1D coupling constant
$g_{\mathrm{1D}}$. Hence, if we tune $g_{\mathrm{1D}}$ near to a CIR either by means of magnetic Feshbach resonances or via accurate control of the transverse trap
frequency $\omega_{\perp}$, we can indeed enhance the atom-phonon coupling up to $(0.1-1)E^*$. Specifically, in order to enhance $g_{\mathrm{1D}}$ by a factor
$10^4-10^5$ we need to tune the ratio $a_{\perp}/\as$ from the CIR position with a precision on the order of $\lesssim10^{-4}$ for $R^*/a_{\perp} = 0.447$ (the corresponding
CIR position is at $a_{\perp}/\as \simeq 1.7424$). Recently, atom-ion Feshbach resonances for the atom-ion pair Li-Yb$^+$ have been investigated~\cite{Tomza:2015}, where
Feshbach resonances are dense at either small magnetic fields (i.e., $\lesssim 150$ G) or at large magnetic fields (i.e., within the range $\sim 500-2000$ G), upon the Ytterbium
isotope (i.e., fermionic vs. bosonic). Since magnetic fields can be controlled with a precision of about 1 mG (see, e.g., Ref.~\cite{Haller:2010,Volz:2003}) and optical trap frequencies can be
controlled with a precision of about 10 mHz, the required precision for tuning the ratio $a_{\perp}/\as$ close to a CIR should be experimentally attainable. We also note that eventually
dressing to a Rydberg state, as recently suggested~\cite{Secker:2016}, would further aid the enhancement of the longitudinal atom-phonon coupling. In this case, however, even though the
dressed atom-ion potential retains the $r^{-4}$ character, the pre-factors are such that the resulting potential is more long-range (on the micrometer range), and therefore a more
detailed analysis of the effect on the CIR position should be undertaken.

In next section we shall investigate numerically the position of the CIR beyond the LWL, and the applicability of the semi-analytic formula (\ref{CIRaic}) at
high colliding energies. The latter will also provide us some indication of the reliance of the CIR position on the temperature for the atom-ion pair $^{6}$Li-$^{171}$Yb$^+$.

\subsection{Beyond the long-wavelength limit: $R^*\gtrsim a_{\perp}$}

In this regime the pseudopotential approximation does not hold anymore and we have to rely on numerical simulations only.
Firstly, we have investigated the CIR position at extremely low energy $E_{\parallel} /E^*=10^{-6}$, that is, in the zero-energy limit, by
integration again the Schr\"odinger equation~(\ref{2Dhamiltonian}) with the boundary condition~(\ref{asymptotics}) for different values of
the ratio $R^*/a_{\perp}$~\footnote{Let us note that in the vertical axis of Fig.~\ref{fig4} the width of the transverse trap $a_{\perp}$ is the same as in the horizontal one. Indeed, we have
first integrated Eq.~(\ref{2Dhamiltonian}) for a fixed value of $a_{\perp}$ and consequently determined the coefficient $C_{12}$ corresponding to the position
of the CIR. Then, we have turned off the external atom trap (i.e., $\omega_{\perp}$) and determined with that $C_{12}$ coefficient the corresponding $\as$. Finally, we
have constructed the ratio $a_{\perp}/\as$ with the previously fixed value of $a_{\perp}$, as we have explained at the beginning of the procedure.}.
Interestingly, we found (see Fig.~\ref{fig4}) a strong dependence of the CIR position on the ratio
$R^*/a_{\perp}$ and an isotope-like effect, i.e. the dependence of
the CIR position on the atom mass: $a_{\perp}/\as =
1.4603 +\Delta(m_A/\mu,R^*/a_{\perp})$.
The last dependence, however, disappears
in the LWL, $R^*/a_{\perp}\rightarrow 0$, where the result of the
confined atom-atom scattering is reproduced: Independently of the
atomic mass, the CIR appears when the s-wave scattering length
$\as$ approaches the value $a_{\perp}/1.4603$ accordingly to
Eq.~(\ref{CIRaa}). Figure~\ref{fig4} also indicates that, in
current experiments, with different atom-ion pairs one can
investigates CIRs in different regions of the parameter
$R^*/a_{\perp}$ (shadowed areas). In this regard, with respect to
atom-atom CIRs, which in current experiments can be studied only
in the ``long wavelength'' region, atom-ion CIRs show a broader
range of interaction regimes, meaning that one has more
flexibility in the tunability of the atom-ion interaction.
Moreover, in the limit $\omega_{\perp}\rightarrow 0$
(i.e., $a_{\perp}=\sqrt{\hbar/(\mu\omega_{\perp})} \rightarrow \infty$),
which corresponds to free-space scattering,
$\as\rightarrow \infty$ (see Fig.~\ref{fig4}a), meaning that the
CIR approaches the s-wave zero-energy resonance in free space,
but the ratio $a_{\perp}/\as$ remains constant and equal to
1.4603. Furthermore, we note that our analysis indicates that the
CIR width increases by enhancing the ratio $R^*/a_{\perp}$ [see
the dependence of the coupling constant $g_{\mathrm{1D}}$ on $a_{\perp}/\as$
presented in Fig.~\ref{fig2}(a)]. This finding is quite interesting,
as it shows that one can enhance the atom-ion interaction via CIRs
much more easily than in the LWL, resulting therefore in an easier
experimental control of the interaction strength.

We have also investigated the CIR position at higher colliding energies [see Fig.~\ref{fig4}(b)]. Figure~\ref{fig4} demonstrates that a strong dependence of  the CIR position on the ratio $R^*/a_{\perp}$
and an isotope-like effect persist with increasing colliding energy. This means that such an effect should be observable also at finite temperature. Furthermore, we observe that the formula~(\ref{CIRaic})
does not hold at $R^*\gtrsim a_{\perp}$, where the CIR position $a_{\perp}/\as(k)$ at the finite energy $E_{\parallel}/E^*=0.117$ begins to exceed the value $a_{\perp}/\as = 1.4603 + \Delta(m_A/\mu, R^*/a_{\perp})$ at zero-colliding energy, which
is already beyond the validity assumptions of Eq.~(\ref{CIRaic}).

The reliance of the atom-ion CIR position on the atom mass crucially depends on the background s-wave atom-ion scattering length $a_{\mathrm{3D}}$.
As we already mentioned in the introduction, the scattering lengths in atom-ion systems have been not yet experimentally measured. Nonetheless, recent theoretical
investigations~\cite{Tomza:2015} based on ab-initio electronic structure calculations have predicted for the Li-Yb$^+$ pair background scattering lengths on the order of 2.2$R^*$
for the singlet and 0.8$R^*$ for the triplet state, whereas $R^*\sim 70$ nm. Hence, such predictions are quite promising also in view of measuring atom-ion CIRs as well as in
observing the aforementioned isotope-like effect. The latter requires definitely less demanding transverse trap frequencies for the atom-ion pair $^{87}$Rb-$^{138}$Ba$^+$. For
instance, for $a_{\perp}\simeq R^*$ the required trap frequency would be $\omega_{\perp} = 2\pi\, 2.2$ kHz, while for $^{23}$Na-$^{171}$Yb$^+$ we have $\omega_{\perp} = 2\pi\, 30$ kHz
and for the pair $^{6}$Li-$^{171}$Yb$^+$ we have $\omega_{\perp} = 2\pi\, 350$ kHz.

\begin{figure}[t!]
\includegraphics[scale=0.35]{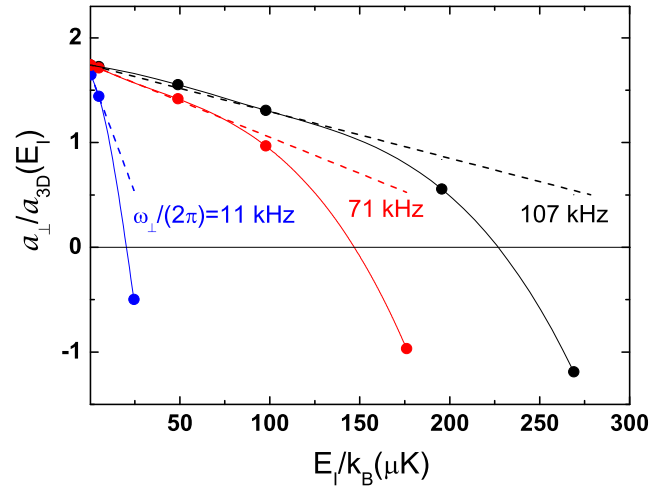}
\caption{ (Color online). The calculated ratio
$a_{\perp}/\as(E_I)$ in the points of the CIR as a function of ion
energy $E_I/k_B$. The calculation has been performed for the
atom-ion pair $^{6}$Li-$^{171}$Yb$^+$ for the three values of the ratio
$R^*/a_{\perp}=(0.173, 0.447, 0.548)$, namely $\omega_{\perp}/(2\pi)
= (11, 71, 107) $kHz. The solid curves link the points
calculated by integrating numerically the 2D Schr\"odinger equation~(\ref{2Dhamiltonian}).
The dashed lines correspond to Eq.~(\ref{CIRaic}).} \label{fig5}
\end{figure}

\section{Impact of micromotion-induced heating}
\label{sec:heating}

As we already pointed out, so far experiments did not yet reach the ultracold s-wave regime because of a fundamental and technical issue named micromotion. Indeed, since ions have to be confined in time dependent fields, they
are not simply at rest in the minimum of a static potential, but they are constantly moving in a potential landscape that can be only approximated as an effective time-independent harmonic confinement. This is particularly true for ions
confined in a so-called Paul trap, where radio-frequency fields are employed to confined the ions in the transverse direction. More precisely, the situation we have in mind can be described by a trap configuration, whose static and
time-dependent electric fields are given by:

\begin{align}
\vec{E}_{\mathrm{s}} &= \frac{m_I}{2\vert e\vert} \omega^2 \left(x,y, -z\right),\nonumber\\
\vec{E}_{\mathrm{rf}} &= \frac{m_I\Omega_{rf}^2 q}{2\vert e\vert} \cos(\Omega_{rf} t)\left(x,-y, 0\right).
\end{align}
Here $\omega$ is the so-called secular frequency, namely the one of the effective harmonic potential [see also Eq.~(\ref{hamiltonian})], $q$ is the dimensionless Mathieu or, alternatively called, instability parameter~\cite{Leibfried:2003},
and $\Omega_{rf}$ the trap drive frequency. By choosing the axis of the waveguide of the colliding atom precisely as the $z$-axis of the Paul trap (see also Fig.~\ref{fig-sketch}) and assuming that the ion is located in the trap centre,
then the effect of the radio-frequency field, and therefore of the micromotion, can be in principle eliminated. Notwithstanding, as it has been shown in Ref.~\cite{Cetina:2012}, when the atom is approaching the ion, the latter will be pulled from
the trap centre because of the polarisation potential~(\ref{interaction}). The displacement of the ion from the trap centre during the collision with the atom can be estimated as~\cite{Cetina:2012}:

\begin{align}
r_c \simeq 1.11\left(\frac{m_A}{m_I}\right)^{5/6} R
\end{align}
with $R=[(R^*)^2\ell_i^4 (m_A+m_I)/(2 m_A)]^{1/6}$ and $\ell_i = \sqrt{\hbar/(m_I\omega)}$. For the atom-ion pair $^{6}$Li-$^{171}$Yb$^+$ we have $r_c \simeq 2.7\,$nm, while for the pair $^{87}$Rb-$^{138}$Ba$^+$ we have $r_c\simeq 21.6\,$nm.
As it can be seen, the effect of the polarisation potential force exerted by the atom on the ion is weaker for the pair $^{6}$Li-$^{171}$Yb$^+$, as the mass ratio $m_A/m_I$ is smaller. As a rough estimate of the corresponding
average micromotion energy of the ion at the collisional point we have: $E_{\mu\mathrm{m}}/k_B= m_I(\omega r_c)^2/(2 k_B) \simeq 0.2\,\mu$K for the pair $^{6}$Li-$^{171}$Yb$^+$ and $E_{\mu\mathrm{m}}/k_B\simeq 9.5\,\mu$K for the pair
$^{87}$Rb-$^{138}$Ba$^+$. We note, however, that these numbers correspond to a rather extreme scenario. Indeed, the effect of the atom in the waveguide on the ion will be significantly smaller, since the ion displacement from the trap centre
will occur mainly along the $z$-axis, where the micromotion is absent. We also note that a recent ab-initio investigation~\cite{Tomza:2015} predicts a ratio of elastic to inelastic scattering cross sections to be larger than 100 for the Li-Yb$^+$ pair
(similar expectations hold for the Rb-Ba$^+$ pair~\cite{Krych:2011}), enabling therefore sympathetic cooling of an Ytterbium ion by Lithium atoms down to temperatures on the order of 10 nK. Given these prospects, at least for this atom-ion pair,
our findings should be indeed observable in a near future both in the LWL and beyond.

In order to be more quantitative, we have computed semi-classically the position of the CIR at higher energies by assuming that the confined atom can be cooled to low temperatures $E_A/k_B= m_A\left<V^2_A\right>/(2k_B)$ -- about a few nK. Since
the corresponding ion energy $E_I/k_B= m_I\left<V^2_I\right>/(2k_B)$ can be on the order of a few tens of $\mu$K, we can reformulate the scattering problem by changing frame of reference, where the atom is moving with velocity -$V_I$ and
colliding with an ion at rest, which is essentially the model described by Eqs.~(\ref{2Dhamiltonian}) and~(\ref{asymptotics}). Hence, the longitudinal energy is given by

\begin{equation}
E_{\parallel} = E_A =\frac{m_A}{m_I}E_I.
\end{equation}
By using this relation we have calculated the dependence of the position of the CIR on the ion energy (i.e., ion temperature) specifically for the atom-ion pair $^{6}$Li-$^{171}$Yb$^+$. Figure~\ref{fig5} shows that at finite colliding energies the CIR position
is well distinguishable upon the transverse trap frequency. It also demonstrates that our semi-analytic formula (\ref{CIRaic}) with the numerically calculated values of $\Delta (m_A/\mu, R^*/a_{\perp})$ (given in Fig.~\ref{fig4}) is valid in a rather broad range
of energies. Thus, at the right side border of Fig.~\ref{fig5}, that is, the black line for which $\omega_{\perp}/(2\pi)=107$ kHz, Eq.~(\ref{CIRaic}) describes indeed the temperature dependence the CIR position with an accuracy better then one percent
in the region up to $E_I/k_B \sim 150 \mu$K. Besides this, Figs.~\ref{fig5} and~\ref{fig3} indicate that by measuring the position of the CIR [i.e., the ratio $a_{\perp}/a_{3D}(E_{\parallel})$ at the point where the CIR appears], the colliding energy $E_{\parallel}$
or the temperature of the confined atomic gas can be determined by the calculated curves $a_{\perp}/a_{3D}(E_{\parallel})$ given in these figures. One can consider this as an alternative way for measuring the temperature of confined either ion or atom(s).

\section{Conclusions}
\label{sec:end}

We have investigated atom-ion CIRs in a wide range of the parameter $R^*/a_{\perp}$ and for different atom-ion pairs.
We have derived  an analytic formula, Eq.~(\ref{CIRai}), for the CIR position in the LWL, which depends on $\as$, $R_0$,
$\alpha_p$, and $a_{\perp}$. We have also provided a semi-analytic formula, Eq.~(\ref{CIRaic}), valid in a wider range
of $R^*/a_{\perp}$. Both results are in very good agreement with our numerical
integration of the 2D Schr\"odinger equation~(\ref{2Dhamiltonian}).

This, at first glance, specific problem about atom-ion CIRs can, however, potentially have a broad range of applications. The obtained
results can be used in near future experiments for searching atom-ion CIRs with the aim of measuring the atom-ion scattering
length, a yet unknown scattering parameter, and manipulating the effective atom-ion interaction in confined systems. In particular, the
control of the atom-ion interaction could be exploited to control the atom-phonon coupling in a solid-state quantum simulator~\cite{Bissbort:2013}
or to investigate more exotic quantum phases in low dimensional systems. In the latter case, the frequency $\omega_{\perp}$ could be tuned in such
a way that simultaneously an atom-atom and an atom-ion CIR are generated and therefore a (hybrid) strongly correlated atom-atom and atom-ion system
is created. Additionally, by exploiting the effect of the complete reflection of the confined atom from the ion in the CIR, one could realise a
device for triggering the confined atom flow, similarly to a single atom transistor~\cite{Micheli:2004}.

Finally, we have qualitatively investigated the impact of the micromotion on the CIR position. We have shown that even though the ion energy is large (a few hundreds of $\mu$K) the effect of confinement-induce
resonances is still discernible, upon transverse atom trap frequency. We note, however, that the impact of this particular ion motion relies on the utilised trapping technology. Indeed, optical traps are significantly
less affected by this problem, although the ion lifetime is on the oder of a few ms~\cite{Huber:2014}. Besides, as we already pointed out previously, recent theoretical investigations~\cite{Cetina:2012,Joger:2014,Krych:2015,Tomza:2015}
have shown that a small atom-ion mass ratio (e.g., Li-Yb$^+$) can indeed help in attaining the s-wave regime and in reducing the impact of the ion micromotion. Hence, our findings are quite general. Instead, it would be
interesting to understand whether CIR can be further controlled via the presence of phonons in the ion trap. This issue as the full quantum mechanical analysis of the ion micromotion on the CIR position will
be investigated in the near future.

\section{Acknowledgements}

The authors are very grateful to R.~Gerritsma and Z.~Idziaszek for a critical reading of the manuscript and for very helpful comments,
and thank P.~Giannakeas, V.~Pupyshev and P.~Schmelcher for discussions. V.~S.~M. acknowledges financial support by the Heisenberg-Landau Program and A.~N. by the
cluster of excellence ``The Hamburg Centre for Ultrafast Imaging" of the Deutsche Forschungsgemeinschaft.

\bibliography{biblio-RG.bib}

\end{document}